\documentclass{article}
\usepackage{spconf,amsmath,graphicx}

\usepackage{enumitem}
\setlist{nosep, leftmargin=14pt}
\let\svthefootnote\thefootnote
\newcommand\freefootnote[1]{%
  \let\thefootnote\relax%
  \footnotetext{#1}%
  \let\thefootnote\svthefootnote%
}
\usepackage[utf8]{inputenc}
\usepackage[english]{babel}
\usepackage{graphicx}
\usepackage{hyperref}

\usepackage{amsfonts} 
\usepackage{float}
\usepackage{xcolor}
\usepackage{caption}
\usepackage{subcaption}
\newcommand{\cmmnt}[1]{\ignorespaces}
\usepackage{eqnarray,amsmath}
\usepackage{amsmath,bm}
\usepackage{amssymb}
\usepackage{graphicx}
\usepackage[linesnumbered,ruled,vlined]{algorithm2e}

\title{Finite Element Reconstruction of stiffness images in MR elastography using statistical physical forward modeling and proximal optimization methods}
\name{%
    Narges Mohammadi$^{\star}$
    \qquad Marvin M. Doyley$^{\star}$%
    \qquad Mujdat Cetin$^{\star}{}^{\dagger}$}
\address{$^{\star}$ Department of Electrical and Computer Engineering, University of Rochester, Rochester, NY, USA \\%
   $^{\dagger}$ Goergen Institute for Data Science, 
University of Rochester, Rochester, NY, USA
}

\begin{document}
\maketitle
\begin{abstract}
Quantitative characterization of tissue properties, known as elasticity imaging, can be cast as solving an ill-posed inverse problem. The finite element methods (FEMs) in magnetic resonance elastography (MRE) imaging are based on solving a constrained optimization problem consisting of a physical forward model and a regularizer as the data-fidelity term and the prior term, respectively. In existing formulation for the elasticity forward model, physical laws that arise from equilibrium equation of harmonic motion, indicate a deterministic relationship between MRE-measured data and unknown elasticity distribution which leads to the poor and unstable elasticity distribution estimation in the presence of noise. Toward this end, we propose an efficient statistical methodology for physical forward model refinement by formulating it as linear algebraic representation with respect to the unknown elasticity distribution and incorporating an analytical noise model. To solve the subsequent total variation regularized optimization task, we benefit from a fixed-point scheme involving proximal gradient methods. 
Simulation results of elasticity reconstruction in various SNR conditions verify the effectiveness of the proposed approach.
\end{abstract}

\section{Introduction}
MRE is an evolving imaging modality with significant potential in clinical diagnosis and tissue characteristic visualization. 
MRE has been successfully used for chronic liver diagnosis as a non-invasive, reliable alternate to liver biopsy and is also being developed for detection of breast, kidneys and lungs cancer malignancy \cite{sinkus2018}. The major profits of elasticity reconstruction using MRE techniques over ultrasound can be described in two folds: first, improved resolution and accuracy can be achieved by MRE measurements as opposed to ultrasound due to its low spatial resolution of lateral displacement\cmmnt{ which is on the basis of acoustic beam width}; second, MRE features enable multi-dimensional displacement measurements. The basic steps of MRE reconstruction can be described as the acquisition of deformation fields called MRE-measured data through an integrated MRI machine and a transducer and then reconstructing the underlying tissue property distribution using this measured data. In this regard, a dynamic external vibration is applied to the top of the soft tissue which leads to internal time-harmonic displacement fields captured by MR imaging techniques. \\
For elasticity imaging, several approaches have been examined based on local frequency estimation (LFE) method, direct inversion method, and indirect FEM-based method \cite{sinkus2018}.
The first two techniques employ a local homogeneity assumption which leads to blurry edges due to the large gradient of elasticity parameters \cite{honarvar}. Moreover, direct inversion methods utilize a  deterministic representation of the equilibrium equation as the physical forward model and estimate the unknown elasticity modulus by linear inversion of this forward model which leads to an unstable solution in noisy conditions. The third one as a model-based indirect approach can be implemented as a regularized optimization problem with improved reconstruction performance without any local homogeneity assumption\cmmnt{ and reduced computational load by using a single frequency}. This constrained optimization problem employs a deterministic physical model of internal deformation pattern and boundary conditions as the forward model which commonly involves a time-harmonic equilibrium condition described as partial differential equations (PDEs).
Existing model-based MRE reconstruction methods\cmmnt{employing governing PDE and physical boundary constraints}, assume an initial elasticity modulus and solve the constrained forward model iteratively until it converges to a stationary solution \cite{marvin}. These approaches result in ill-conditioned problems, leading to poor solutions in low SNR settings and expensive computation time \cite{MRE-AI} . \\To tackle these shortcomings of MRE model-based elastography, we propose a new statistical \cmmnt{iterative} algorithm for estimating elasticity distributions in the presence of noise. In this approach, a refined objective function is developed by integrating linear algebraic modeling of PDE conditions and analytical error modeling of elasticity parameters leading to a unified physical forward model. Moreover, the proposed objective function is developed by augmenting \cmmnt{the principled statistical forward model by} total variation (TV) regularization for preserving sharp elasticity transitions at the edges. This optimization problem is iteratively solved using fixed-point algorithms and proximal gradient methods.  \cmmnt{For simplicity of theoretical analysis, MRE-measurements are conducted in 2D \cite{2DMRE} which would be extended to 3D by parallel computation as well.} Our simulation results verify the effectiveness of the proposed methodology.\\
The rest of this paper is organized as follows. In Section \ref{sec2}, we analyze the MRE forward model to achieve a unified linear representation of the governing PDEs. The MRE inverse problem and the proposed paradigm as its solver are elaborated in Section \ref{sec3}. The simulation results \cmmnt{of elasticity image reconstruction using synthetic MRE measurements} are presented in Section \ref{sec4}, and finally, concluding remarks are provided in Section \ref{sec5}. 

\section{Forward Problem Statement}
\label{sec2}
\cmmnt{Mechanical waves in a elastic medium may appear due to several reasons. Time-varying forces on the boundary of the material are one class of possible source. Likewise, time-varying body forces in the interior can generate a wave, or a disturbance propagate through the undisturbed medium.}
In the MRE imaging problem, the harmonic equation of motion is described by PDEs known as equilibrium conditions which relate measured displacements with unknown elasticity parameters of the tissue. Utilizing an irregular triangle mesh for cross-section discretization of the tissue over the nodes, we aim to put forward a compact linear representation for the discretized PDEs which requires a detailed understandings of them in node, element, and mesh extents.

\subsection{Node Analysis}
\cmmnt{For elastography application, it is required to estimate the elasticity distribution of material given motion (displacement) measurements.}
The governing PDE of harmonic motion in an isotropic linear elastic medium \cmmnt{\cite{curl3}} for each node can be represented as:
\begin{equation}
\label{eq:1}
    \nabla\cdot\left [ \mu\left ( \nabla\Bar{\mathbf{q}}+(\nabla\Bar{\mathbf{q}})^{T} \right ) +\lambda (\nabla\cdot\Bar{\mathbf{q}})\right ]=\rho \frac{\partial^2 \Bar{\mathbf{q}}}{\partial t^2}
\end{equation}
where $\Bar{\mathbf{q}}\in \mathbb{R}^{2\times 1}$ is the displacement vector in time domain consisting of the lateral and the axial displacement of each node, $\lambda$ and $\mu$ denote the Lame parameters, and $\rho$ is the tissue density.
The linear elastic wave equation for isotropic tissues in frequency domain would be described as: 
\begin{equation}
\label{eq:2}
\left [ \mu(q_{i,j}+ q_{j,i}) \right ]_{,j}+(\lambda q_{j,j})_{i}=-\rho\omega^{2}q_{i} 
\end{equation}
where $i,j$ refer to Cartesian axes and indices after comma denote differentiation ($q_{i,j}=\frac{\partial^2 q_{i}}{\partial x_{j}}$), $q\in \mathbb{R}^{2\times 1}$ represents the Fourier displacement field, and $\omega$ is the stimulator frequency. When we have a linear elastic and isotropic medium, $\lambda$ and $\mu$ becomes two scalar unknown parameters instead of a function of the position, and (\ref{eq:2}) can be formulated as an algebraic matrix equation. To this end\cmmnt{To satisfy homogeneity condition, finite element discretization is implemented which assumes constant $\lambda$ and $\mu$ in each element of the mesh. In this case}, the local equilibrium equation for each node could be rewritten using \cmmnt{\cite{MRE_elsticity}, \cite{marvin}} \cite{marvin},\cite{MRE_book} as: 
\begin{equation}
\label{eq:5}
\mu q_{i,j,j}+ (\lambda+\mu) q_{j,j,i}=-\rho\omega^{2}q_{i} 
\end{equation}
and these equations can be solved separately at each node using only data from a local region to estimate local derivatives \cite{AIDE}, \cite{curl2017}\cmmnt{\cite{MRE_elsticity}, ,}. To have a linear algebraic representation of the PDE in (\ref{eq:5}), the following nodal model is introduced in \cite{AIDE} as:
\begin{equation}
\label{eq:7}
A\begin{bmatrix}
\lambda+\mu\\ 
\mu
\end{bmatrix}=-\rho\omega^{2}\begin{bmatrix}
q_{i}\\ 
q_{j}
\end{bmatrix}
\qquad A=\begin{bmatrix}
q_{j,j,i}&q_{i,j,j}\\ 
q_{i,i,j}&q_{j,i,i}
\end{bmatrix}
\end{equation}

\subsection{Element Analysis}
\label{element}
To solve these equations for each element of the discretized medium, we define the differentiation operator $B$ (where $B\in \mathbb{R}^{3\times 6}$ for a 2D triangular element) as the generalized strain-displacement transformation matrix as follows:
\begin{equation}
\label{eq:9}
\begin{array}{l}
\mathbf{B}=\frac{1}{2\Delta}[B_{1} B_{2} ... B_{M}] \qquad
B_{m}=\begin{bmatrix}
\frac{\partial }{\partial x} &0 \\ 
 0&\frac{\partial }{\partial y}  \\ 
 \frac{\partial }{\partial y} & \frac{\partial }{\partial x}  
\end{bmatrix}
\end{array}
\end{equation}
where $M$ is the number of nodes in each element and $\Delta$ is the element area. The harmonic equilibrium equation of each element can be described with a linear algebraic model as follows:
\begin{equation}
\label{eq:10}
\mathbf{B}^{T}\mathbf{C}\mathbf{B}\mathbf{q}_{e}=-\rho\omega^{2}\mathbf{q}_{e}
\end{equation}
where $\mathbf{q}_{e}\in \mathbb{R}^{6\times1}$ consists of lateral and axial Fourier displacement fields of the three nodes of each element \cmmnt{\cite{MRE_elsticity}} and $\mathbf{C}$ is the stress-strain matrix defined as:
\begin{equation}
\label{eq:11}
\mathbf{C}=\begin{bmatrix}
\lambda+2\mu &\lambda &0 \\ 
 \lambda & \lambda+2\mu &0  \\ 
 0&0&\mu 
\end{bmatrix}=E\Tilde{\mathbf{C}}
\end{equation}
\begin{equation}
\label{eq:12}
\begin{array}{l}
\lambda=\frac{\nu E}{(1+\nu)(1-2\nu)}, \quad \mu=\frac{E}{2(1+\nu)} \qquad \text{for plain strain}\\

\end{array}
\end{equation}
Here, $E$ is element elasticity modulus as a scalar value and $\nu$ is the Poisson's ratio. To extend (\ref{eq:10}) to any point inside the element $\mathbf{q}_{e}(x)$, we define shape function $\Phi$ as used in \cite{vanhouten} to interpolate $\mathbf{q}_{e}(x)$ using its nodal displacement values $\mathbf{u}_{e}$ by $\mathbf{q}_{e}(x)=\Phi\mathbf{u}_{e}$ which leads to the local equilibrium equation as follows:
\begin{equation}
\label{eq:13}
\mathbf{B}^{T} \mathbf{C} \mathbf{B} \Phi \mathbf{u}_{e}=-\rho\omega^{2}\Phi \mathbf{u}_{e}
\end{equation}
For solving the aforementioned equation, Galerkin method proposes residual minimization by multiplying both sides of (\ref{eq:13})  by the shape function, integrating over the element and equating to zero:
\begin{equation}
\label{eq:14}
\int_{V}\Phi^{T}\mathbf{B}^{T}E\Tilde{\mathbf{C}}\mathbf{B}\Phi dv \mathbf{u}_{e} +\int_{V}\rho\omega^{2}\Phi^{T}\Phi dv \mathbf{u}_{e}=0
\end{equation}

For more compact representation of (\ref{eq:14}), let us define the following variables:
\begin{equation}
\begin{array}{l}
\label{eq:15}
\mathbf{k}_{e}(E)=[\bm{\psi}^{T} E]=\int_{V}\Phi^{T}\mathbf{B}^{T}E\Tilde{\mathbf{C}}\mathbf{B}\Phi dv\\
\\
\bm{\psi}^{T}=\int_{V}\Phi^{T}\mathbf{B}^{T}\Tilde{\mathbf{C}}\mathbf{B}\Phi dv
\end{array}
\end{equation}
\begin{equation}
\label{eq:16}
\mathbf{k}'_{e}=\int_{V}\rho\omega^{2}\Phi^{T}\Phi dv 
\end{equation}
Using this notation and incorporating  $\mathbf{f}_{e}$ as the force boundary conditions (BCs), local equilibrium equation for each element \cmmnt{\cite{harmonic2000}} could be expressed as: 
\begin{equation}
\label{eq:17}
(\mathbf{k}_{e}(E)+\mathbf{k}'_{e})\mathbf{u}_{e}=\mathbf{f}_{e}
\end{equation}
where $\mathbf{k}_{e}(E)\in\mathbb{R}^{2M\times2M}$, $\mathbf{k}'_{e}\in\mathbb{R}^{2M\times2M}$, $\mathbf{u}_{e}\in\mathbb{R}^{2M\times 1}$, $\mathbf{f}_{e}\in\mathbb{R}^{2M\times 1}$ and (\ref{eq:17}) is called the local stiffness equation.
\subsection{Mesh Analysis}
By assembling the local equilibrium equation of all elements of the mesh, \cmmnt{each nodal vector is concatenated in a global vector and each local matrix is transformed to the global one; therefore, }the global equilibrium equation could be introduced \cmmnt{\cite{asilomar}} as:
\begin{equation}
\begin{array}{l}
\label{eq:18}
\mathbf{K(E)}\mathbf{u}=(\mathbf{\Psi}^{T}\mathbf{E})\mathbf{u}=\mathbf{f_{true}}\\
\mathbf{D(u)}\mathbf{E}=(\mathbf{\Psi}\mathbf{u})^{T}\mathbf{E}=\mathbf{f_{true}}
\end{array}
\end{equation}
If $N$ denotes the number of nodes in the mesh, then $\mathbf{K(E)}\in \mathbb{R}^{2N\times2N}$, $\mathbf{D(u)}\in \mathbb{R}^{2N\times2N}$, $\mathbf{u}\in \mathbb{R}^{2N\times1}$, $\mathbf{E}\in \mathbb{R}^{N\times1}$ , $\mathbf{\Psi}\in \mathbb{R}^{N\times2N\times2N}$ and  $\mathbf{f_{true}}\in \mathbb{R}^{2N\times1}$ which is applied as Neumann BC on measured Fourier displacement vector.

\section{Inverse Optimization Problem formulation}
\label{sec3}
The statistical representation of the MRE forward model which reveals the relationship between tissue elasticity parameters and internal deformation data can be described as:
\begin{equation}
\label{eq:19-1}
\mathbf{f}=\mathbf{D(u)}\mathbf{E}+\mathbf{w}\qquad \mathbf{w}\sim \mathcal{N}(0,\,\bm{\Sigma_{w}})
\end{equation}
where $\mathbf{f}$ contains the observed force BCs and $\mathbf{w}\in \mathbb{R}^{2N\times 1}$ is the \cmmnt{nodal} Gaussian noise vector. 
The frequency domain displacement fields are obtained using Fourier transform of phase contrast imaging which introduce the observation model $
\mathbf{u^{m}}=\mathbf{u}+\mathbf{n}$ where $\mathbf{n}\sim \mathcal{N}(0,\,\bm{\Sigma_{n}})$ and $\mathbf{u^{m}}$ is the contaminated Fourier displacement fields with noise $\mathbf{n}\in \mathbb{R}^{2N\times 1}$ with covariance $\bm{\Sigma_{n}}$ which can capture noise variance in the lateral and axial direction.
Merging \cmmnt{fusing=unifying} the statistical forward model in (\ref{eq:19-1}) with the displacement observation model yields to: 
\begin{eqnarray}
\label{eq:19-3}
\mathbf{f}&=&\mathbf{K}(\mathbf{E})\mathbf{u}+\mathbf{w}=\mathbf{K}(\mathbf{E})(\mathbf{u^{m}}-\mathbf{n})+\mathbf{w}\nonumber\\
&=&\mathbf{K}(\mathbf{E})\mathbf{u^{m}}-\mathbf{K}(\mathbf{E})\mathbf{n}+\mathbf{w}
\end{eqnarray}
Setting ${\mathbf{\Tilde{w}}}=-\mathbf{K}(\mathbf{E})\mathbf{n}+\mathbf{w}$ and utilizing $\mathbf{D}(\mathbf{u^{m}})\mathbf{E}=\mathbf{K}(\mathbf{E})\mathbf{u^{m}}$ and plugging these in (\ref{eq:19-1}) leads to the following joint observation model:
\begin{equation}
\label{eq:19-4}
\mathbf{f}=\mathbf{D}(\mathbf{u^{m}})\mathbf{E}+\mathbf{\Tilde{w}}\qquad \mathbf{\Tilde{w}}\sim \mathcal{N}(0,\,\bm{\Gamma})
\end{equation}
where $\bm{\Gamma}$ is defined by:
\begin{equation}
\label{eq:19-5}
\bm{\Gamma}=\bm{\Sigma_{w}}+\mathbf{K}(\mathbf{E})\bm{\Sigma_{n}}\mathbf{K}(\mathbf{E})^{T}
\end{equation}
Hence our joint observation model in  (\ref{eq:19-4}) can be interpreted as involving signal dependent correlated noise. By having $\mathbf{f}$ and $\mathbf{u^{m}}$ measurements, it is required to solve a regularized optimization problem to estimate the unknown elasticity modulus $\mathbf{E}$. For achieving a stable maximum {\it a posteriori} (MAP) estimation, we develop a TV-constrained cost function as: 
\begin{equation}
\label{eq:20}
\begin{array}{l}
\mathbf{\hat{E}}=\mathrm{argmin} _{\mathbf{E}}\quad\frac{1}{2}\left \|  \mathbf{f}-\mathbf{D}(\mathbf{u^{m}})\mathbf{E} \right \|_{{\bm{\Gamma}}^{-1}}^{2}+\frac{N}{2}\mathrm{log}\left | \bm{\Gamma} \right |+\lambda \|\nabla\mathbf{E}\|_{1}\\
\quad\quad\quad
s.t.\quad \mathbf{E}>0
\end{array}
\end{equation}
where $\left \| \mathbf{A} \right\|_{\mathbf{B}}^{2}:=(\mathbf{A}^{T}\mathbf{B}\mathbf{A})$. For solving (\ref{eq:20}), a fixed-point method \cite{fixedpoint} is established by fixing $\bm{\Gamma}$ while we update $\mathbf{E}$, and then this new $\mathbf{E}$ is fed into (\ref{eq:19-5}) to update $\bm{\Gamma}$.
We exploit proximal gradient methods \cite{proximalnew} for updating $\mathbf{E}$ \cmmnt{in each step of the fixed-point approach} as follows \cite{asilomar} :
\begin{equation}
\label{eq:22}
\mathbf{E}_{n+1}=\textrm{prox}_{\mathbf{E}_{n}>0}(\textrm{prox}_{\gamma _{n}TV}(\mathbf{E}_{n}-\gamma _{n}\nabla g(\mathbf{E}_{n})))
\end{equation}
\begin{equation}
\label{eq:23}
\begin{array}{l}
g(\mathbf{E})=\frac{1}{2}(\mathbf{f}-\mathbf{D}(\mathbf{u^{m}})\mathbf{E}  )^{T}\bm{\Gamma}^{-1}(\mathbf{f}-\mathbf{D}(\mathbf{u^{m}})\mathbf{E}  )\\
\nabla g(\mathbf{E})=-(\mathbf{D}(\mathbf{u^{m}}))^{T}\bm{\Gamma}^{-1}(\mathbf{f}-\mathbf{D}(\mathbf{u^{m}})\mathbf{E} )
\end{array}
\end{equation}
\section{Simulations and Results}
\label{sec4}
For evaluating the performance of the proposed elastography method, we aim to reconstruct the elasticity modulus $\mathbf{E}$ utilizing the noisy Fourier displacement measurements $\mathbf{u^{m}}$ which are also called phase difference fields and the noisy synthetic measurements of force  $\mathbf{f}$ employed as Neumann BCs. Irregular triangle elements are established using FEA for medium discretization over the nodes leading to MRE measurements of dimension $2N\times1$ to represent lateral and axial measurements of mesh nodes. Synthetic clean phase difference fields $\mathbf{u}$ are generated by solving the deterministic forward model $\mathbf{K(E)}\mathbf{u}-\mathbf{f}=0$ for a medium with known elasticity modulus $\mathbf{E}$. Noisy phase difference measurements are obtained by adding multivariate Gaussian noise with covariance $\bm{\Sigma_{n}}$ and noise level $\Delta =\left \| \mathbf{u^{m}}-\mathbf{u} \right \| / \left \| \mathbf{u^{m}} \right \|$ between $0.1-20\%$.
The transducer stimulus frequency (typically 20-200Hz) is set to $\omega=90Hz$, the tissue density is $\rho=1000Kg/m^{3}$ for soft tissues which are mostly composed of water, and Poisson's ratio $\nu$ is set to 0.495.
For elasticity reconstruction, the optimization problem described in Section \ref{sec3} is solved using fixed point and proximal gradient methods. We compared our proposed approach with OpenQSEI \cite{matlab}, as one of the conventional iterative approaches which employ deterministic representation of harmonic motion PDEs. It is worth mentioning that the applied global stiffness matrix in OpenQSEI is modified to match the procedure introduced in Section \ref{element} to account for harmonic motion instead of quasi-static motion. Reconstructed elasticity images by both approaches for different SNRs presented in Fig. \ref{fig:fig1} indicate the proposed method significantly improves the reconstruction performance especially in low SNR.
\cmmnt{Synthetic Fourier-domain noisy displacement $\mathbf{u}^{m}$ (phase image) with $SNR=42dB$ is depicted in Fig. 1}
To perform a quantitative comparison, two performance metrics namely, CNR (contrast-to-noise ratio) and RMS (relative mean square) error are depicted in Fig. \ref{fig:3} which illustrate the superior performance of the proposed approach compared to OpenQSEI with two different regularizers: TV and weighted-smoothness (ws). The Python code of this implementation is available at GitHub \footnote{\url{https://github.com/narges-mhm/MRE-elast}}.
\cmmnt{
\begin{figure}[htb]
\hspace{5em}
\begin{minipage}[b]{0.24\linewidth}
  \centering
  \centerline{\includegraphics[width=6cm]{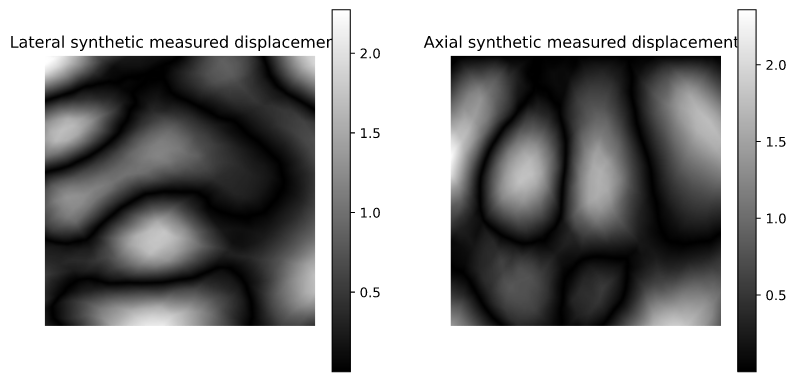}}
  \centerline{(a) \scriptsize{lateral phase image }}\medskip
\end{minipage}
\hspace{7em}
\begin{minipage}[b]{0.24\linewidth}
  \centering
  \centerline{\includegraphics[width=3cm]{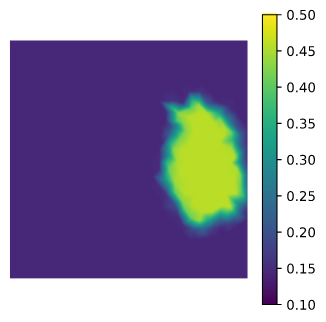}}
  \centerline{(b) \scriptsize{axial phase image }}\medskip
\end{minipage}
\caption{Ground-truth elasticity modulus image. The elasticity of inclusion is 0.46 and for the background is 0.145. The unit of colorbar is 100KPa}
\label{fig:1-2}
\end{figure}
}
\begin{figure}[t]
\begin{minipage}[b]{0.2\linewidth}
  \centering
  \centerline{\includegraphics[width=2.1cm]{groungtruthE.png}}
  \vspace{-0.5\baselineskip}
  \centerline{\scriptsize{(a)} \tiny{Ground truth}}\medskip
\end{minipage}
\hspace{0.5em}
\begin{minipage}[b]{0.2\linewidth}
  \centering
  \centerline{\includegraphics[width=2.5cm]{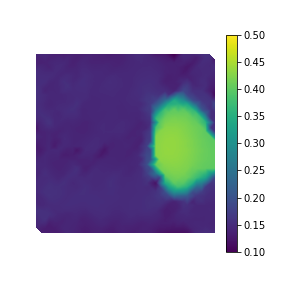}}
  \vspace{-1\baselineskip}
  \centerline{\scriptsize{(b)} \tiny{OpenQSEI, SNR=42dB  }}\medskip
\end{minipage}
\hspace{1em}
\begin{minipage}[b]{0.2\linewidth}
  \centering
  \centerline{\includegraphics[width=2.5cm]{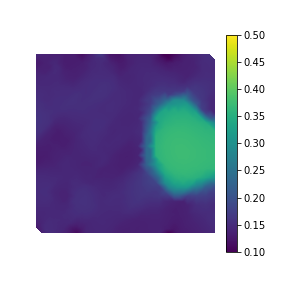}}
  \vspace{-1\baselineskip}
  \centerline{\scriptsize{(c)} \tiny{OpenQSEI, SNR=35dB  }}\medskip
\end{minipage}
\hspace{1em}
\begin{minipage}[b]{0.2\linewidth}
  \centering
  \centerline{\includegraphics[width=2.5cm]{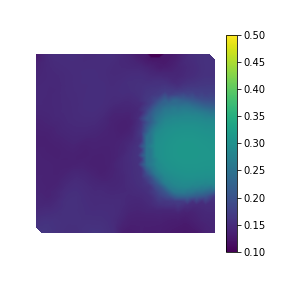}}
  \vspace{-1\baselineskip}
  \centerline{\scriptsize{(d)} \tiny{OpenQSEI, SNR=25dB }}\medskip
\end{minipage}\par\vspace{-1\baselineskip}
\begin{minipage}[b]{0.16\linewidth}
  \centering
  \centerline{\includegraphics[width=2.2cm]{blank.png}}
  \vspace{-1\baselineskip}
  \centerline{}\medskip
\end{minipage}
\hspace{2em}
\begin{minipage}[b]{0.16\linewidth}
  \centering
  \centerline{\includegraphics[width=2.4cm]{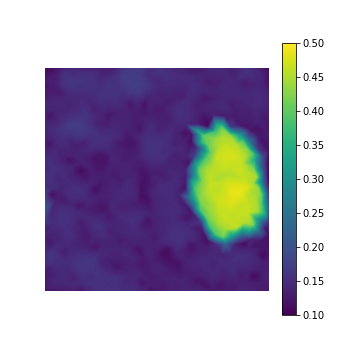}}
  \vspace{-1\baselineskip}
  \centerline{\scriptsize{(e)} \tiny{Proposed, SNR=42dB }}\medskip
\end{minipage}
\hspace{2em}
\begin{minipage}[b]{0.16\linewidth}
  \centering
  \centerline{\includegraphics[width=2.4cm]{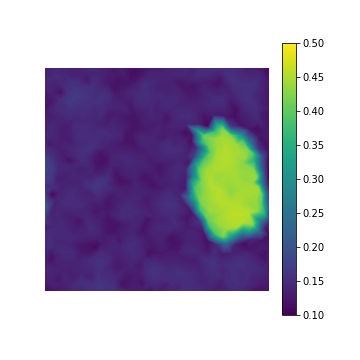}}
  \vspace{-1\baselineskip}
  \centerline{\scriptsize{(f)} \tiny{Proposed, SNR=35dB }}\medskip
\end{minipage}
\hspace{2em}
\begin{minipage}[b]{0.16\linewidth}
  \centering
  \centerline{\includegraphics[width=2.4cm]{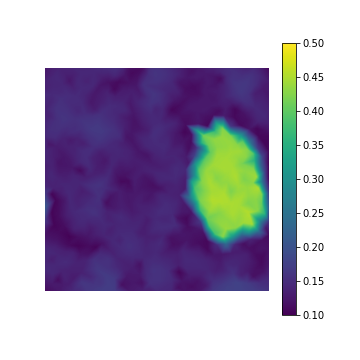}}
  \vspace{-1\baselineskip}
  \centerline{\scriptsize{(g}) \tiny{Proposed, SNR=25dB }}\medskip
\end{minipage}
\vspace{-0.5\baselineskip}
\caption{Ground truth and reconstructed elasticity modulus \cmmnt{$\hat{\mathbf{E}}$} with TV regularization for three different SNRs. $\mathbf{E}_{true}=0.46 $ for the inclusion and $\mathbf{E}_{true}=0.145$ for the background. The unit of the color bar is 100 KPa.}
\label{fig:fig1}
\end{figure}

\begin{figure}[ht]
 \centering
\includegraphics[width=7cm]{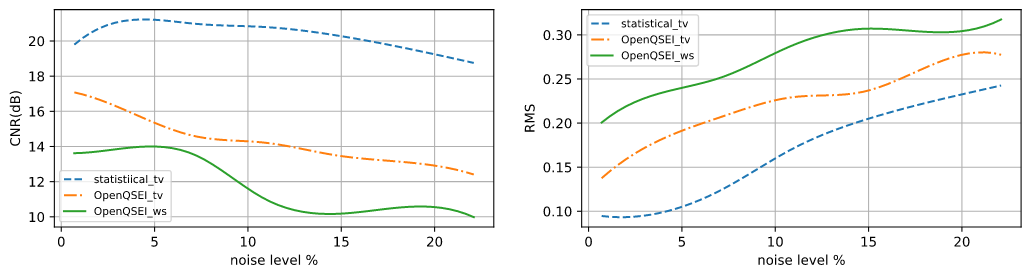}
\caption{CNR and RMS performance metrics for noise levels $\Delta=0.1-20\%$ achieved by the proposed approach and OpenQSEI with TV and weighted-smoothness (ws) regularizers.}
\label{fig:3}
\end{figure}
\section{Conclusion}
\label{sec5}
This article proposes a new statistical imaging methodology for model-based MRE by solving a constrained inverse problem. Proposed framework involves a unified objective function embedding a linear algebraic forward model of the governing physical PDE and a total variation regularizer. The physical forward model incorporates statistical models of noise involved in force and Fourier displacement measurements, which leads to a signal dependent correlated noise modeling. We utilize a fixed-point iterative approach for solving the elasticity optimization problem which is built on proximal gradient algorithms. The propose approach is a basis for 3D MRE reconstruction. The simulation results demonstrate the effectiveness of the proposed approach, even in the case of severe noisy measurement fields.

\section{Compliance with Ethical Standards}
\label{sec:ethics}
This is a numerical simulation study for which no ethical approval was required.
\section{Acknowledgments}
\label{sec:ack}
This work has been partially supported by the National Science Foundation (NSF) under Grant CCF-1934962.

\bibliographystyle{IEEEbib}
\bibliography{M335}

\end{document}